\documentstyle[12pt,cite,epsfig]{article}

\begin{document}

\author{C.Pigorsch and S.Trimper \\
Martin-Luther-Universit\"{a}t Halle-Wittenberg,\\
Fachbereich Physik, 06099 Halle Germany}
\title{An Extended Network Model with a Packages Diffusion Process}
\date{\today }
\maketitle

\begin{abstract}
\noindent The dynamics of a packages diffusion process within a
selforganized network is analytically studied by means of an extended $f$%
-spin facilitated kinetic Ising model (Fredrickson-Andersen model) using a
Fock-space representation for the master equation. To map the three
component system (active, passive and packages cells) onto a lattice we
apply two types of second quantized operators. The active cells correspond
to mobile states whereas the passive cells correspond to immobile states of
the Fredrickson-Andersen model. An inherent cooperativity is included
assuming that the local dynamics and subsequently the local mobilities are
restricted by the occupation of neighboring cells. Depending on a
temperature-like parameter $h^{-1}$ (interconnectivity) the diffusive
process of the packages (information) can be almost stopped, thus we get a
well separation of the time regimes and a quasi-localization for the
intermediate range at low temperatures.
\end{abstract}

\section{Introduction}

During the last years there is a huge effort to understand the kinetics of
non-equilibrium phenomena. A wide range of discrete and continuous models
for such processes is analytically and numerically studied. The problems of
interest in this context concern the crystal growth, transport (traffic)
models, diffusion processes and supercooled liquids \cite
{Godreche,Privman,Matthis,Schuetz0,Schulz0,Pigorsch0}.

Here, we will apply the kinetics of the Fredrickson-Andersen model (FAM)
recently discussed in the framework of the glass transition and related
phenomena \cite
{Schulz0,Pigorsch0,Fredrickson1,Fredrickson2,Fredrickson3,Butler,Schulz1,Heuer}%
. But we show that this model may be used on other fields like stock
trading, citation networks, company relations or internet communications as
well \cite{Stauffer0,Lawrence,Kirman,Redner}. In general, we study the
diffusion of information within a network system of active links and
passive/active cells (or nodes). The switch between a passive and active
cell is controlled by a temperature-like parameter $h^{-1}$which can be
interpreted as the interconnectivity. At the maximum interconnectivity (at
infinite temperature) the system possesses equal parts of active and passive
cells (nodes) whereas at the minimum interconnectivity (at zero temperature)
there are only passive cells (nodes). Furthermore, the alteration is also
controlled by the nearest neighborhood. If enough adjacent active cells
(nodes) exist (more than a fixed number $f$) the active cell (node) can
become passive and vice versa. Thus, only a sufficiently active environment
may determine and alter the state of a cell (node) like in a citation
community where only accepted (active) people may decide about the worth of
an opinion of a member in a related field. To this network formed by
passive/active cells (nodes)\ and active links, consisting of two adjacent
active cells (nodes), we add further particles which may be assigned to
(information) packages. These particles can only diffuse along active links
but are confined by passive cells (nodes). Therefore, we have a diffusion in
a self-organized network where information may stick (and therefore be
localized) at passive cells (nodes) or runs through a network of active
links. This is the same situation like in the internet where passive routers
cannot transfer any data but data on active router should be directed to the
next active router (Of course, the difference is that this is a directed
motion whereas we consider a non-directed motion here.).

Now we demonstrate how one may relate this network to the FAM originally
dedicated to supercooled liquids. For such systems one assumes that the
dynamics can be roughly regarded as independent of the microscopic details,
thus the analysis is based on a mesoscopic formulation. Hence, one reduces
the degrees of freedom to a smaller set of relevant observables following
the original idea of Fredrickson and Andersen \cite
{Fredrickson1,Fredrickson2,Fredrickson3}. The supercooled liquid as the
many-body system is treated on a virtual lattice with sufficiently small
cells of size $l$ where the lattice should not have any significant
influence on the underlying dynamics. However, in case of our network model
we will immediately start on the mesoscopic scale (e.g. on the elementary
scale of routers where the macroscopic scale is the total internet).

The cell structure enables us to attach to each cell $j$ a two-state
observable $\sigma _{j}$ which characterizes its actual state. The
realization is given by the local activity $m_{j}$ with $\sigma _{j}=-1$
(spin down) if $m_{j}<\bar{m}$ (passive state) and $\sigma _{j}=+1$ (spin
up) if $m_{j}>\bar{m}$ (active state) where $\bar{m}$ is a fixed threshold
of the system. The threshold is chosen in this way that all cells are
passive at zero temperature. Active (passive) states may be also denoted as
mobile (immobile) states. The number of cells in the active and the passive
state, respectively, have not to be conserved. (Notice that in case of
supercooled liquids the usual separation is done by two different density
states although this is more connected to a Kawasaki exchange dynamics
rather than to a Glauber flip dynamics\cite{Fredrickson3}. This mapping
implies different densities of such cells; $\sigma _{j}=-1$ corresponds to
the more solid-like state and $\sigma _{j}=1$ corresponds to the more
liquid-like state.) Thus, the real kinetics is more Glauber-like
(non-conserved) dynamics thus we suppose that the basic dynamics is a flip
process $\sigma _{j}=+1\leftrightarrow \sigma _{j}=-1$. It is controlled by
self-induced topological restrictions introduced below. This type of
dynamics leads to a relaxation behavior resembled to that of a cooperative
system. In particular, an elementary switch at a given cell is allowed only
if the number of nearest neighbor mobile/active cells ($\sigma _{j}=+1$) is
equal to or larger than a certain restrictive number $f$ with $0\leq f\leq z$
($z$ is the coordination number of the lattice). Hence, elementary flip
processes combined with the restrictions may lead to the cooperative
arrangements within the underlying network system.

As demonstrated such a mesoscopic approach is able to model supercooled
liquids as well. The Fredrickson-Andersen model (or $f$--spin facilitated
kinetic Ising model) \cite{Fredrickson1,Fredrickson2,Fredrickson3} was
analytically \cite{Schulz0,Pigorsch0,Schulz2} and numerically \cite
{Butler,Schulz1,Heuer,Schulz3} studied in broad variety. The FAM can be
classified as an Ising-like model of which kinetics is confined by
restrictions of the ordering of nearest neighbors to a given lattice cell.
These self--adapting environments especially influence the long-time
behavior of the spin-relaxation\cite{Schulz1,Pigorsch0} particularly
relevant for supercooled liquids.

Additionally, here we add particles to the system of which diffusive
dynamics are coupled with the existence of the active states. Therefore, we
require that active cells are necessary for the motion of packages because
passive states block any packages diffusion. Or in other words data are
fixed and static as long as they are assigned to passive cells, but
information may diffuse through a network of active cells. In the present
work we will incorporate this features into the Fock-space representation of
the master equation in order to compute the packages concentration in a
continuous mean-field-like approximation.

\section{Algebraic Representation and Master Equation}

\noindent Firstly, we give a short survey about the Fock-space method (known
as quantum Hamiltonian as well - more details can be found in e.g.\cite
{Matthis}) A certain state in our lattice system can be characterized by a
set of discrete numbers $\vec{n}=\{n_{i}\}$ (or resp. $\vec{v}=\{v_{i}\}$)
where $n_{i},v_{i}\in \left\{ 0,1\right\} $ denote the local state of a
lattice cell $i$. Furthermore, the following convention is used: $n_{i}=1$ ($%
0$) is related to the active (passive) state, also denoted by $A$ ($B$). The
state $v_{i}=1$ ($0$) corresponds to a cell $i$ occupied (non-occupied) by a
package. We start from the one-step master equation formulated in the
following form 
\begin{equation}
\partial _{t}P(\vec{n},\vec{v},t)=L^{\prime }P(\vec{n},\vec{v},t)
\label{master-equation}
\end{equation}
where $P(\vec{n},\vec{v},t)$ is the probability for a certain configuration $%
\left\{ \vec{n},\vec{v}\right\} $. The linear operator $L^{\prime }$
specified by the dynamics of the system describes the time evolution. Then,
following \cite{Doi1,Doi2}, the probabilities $P(\vec{n},\vec{v},t)$ can be
related to the Fock-space state vector $\mid F(t)\rangle $ as a weight using
the decomposition into the base vectors $\mid \vec{n}\rangle \otimes \mid 
\vec{v}\rangle $ of a orthonormal vector space,
\begin{equation}
\mid F(t)\rangle =\sum_{\vec{n},\vec{v}}P(\vec{n},\vec{v},t)\mid \vec{n}%
\rangle \otimes \mid \vec{v}\rangle \mbox{.}
\end{equation}
This equation enables the quantum formulation for the master equation
resulting in an equivalent manner 
\begin{equation}
\partial _{t}\mid F(t)\rangle =\hat{L}\mid F(t)\rangle 
\label{Time-evolution}
\end{equation}
where $L^{\prime }$ is assigned to the operator $\hat{L}$ in the Fock-space
representation. The procedure had been originally derived for Bose-like
system \cite{Doi1,Doi2} and was later applied also to Fermi-like systems 
\cite{Schuetz0,Sandow,Schuetz1}. Recently, we have proposed a further
extension applying Para-Fermi statistics for different restricted occupation
numbers \cite{Pigorsch0,Schulz4}. The average of a physical quantity $G(\vec{%
n},\vec{v})$ is given by the trace over $\hat{G}$: 
\begin{equation}
\langle \hat{G}(t)\rangle =\sum_{\vec{n},\vec{v}}P(\vec{n},\vec{v},t)G(\vec{n%
},\vec{v})=\langle \vec{r}\mid \hat{G}\mid F(t)\rangle   \label{Average}
\end{equation}
where $\langle \vec{r}\mid $ is the reference state related to the base
vectors by 
\begin{equation}
\langle \vec{r}\mid =\sum_{\vec{n},\vec{v}}\langle \vec{n}\mid \otimes
\langle \vec{v}\mid =\bigotimes \left( 
\begin{array}{c}
1 \\ 
1
\end{array}
\right) \mbox{.}  \label{reference-state}
\end{equation}
The reference state is completely determined by the base $\left\{ \mid \vec{n%
}\rangle ,\mid \vec{v}\rangle \right\} $ of the Fock space and not assigned
to the special model or to the evolution operator $\hat{L}$. The
conservation of the total probability is manifested by $\langle \vec{r}\mid 
\hat{L}=0$. Therefore, the equation of motion is given by 
\begin{equation}
\partial _{t}\langle \hat{G}(t)\rangle =\langle \vec{r}\mid \hat{G}\hat{L}%
\mid F(t)\rangle =\langle \vec{r}\mid \left[ \hat{G},\hat{L}\right]
_{\_}\mid F(t)\rangle \mbox{.}  \label{Equation-of-Motion}
\end{equation}
Notice that the dynamical equation depends on both the algebraic properties
of the underlying operators and the mathematical structure of $\hat{L}$.
Thus, we introduce the second quantized lowering $a_{i}$($v_{i}$) and
raising $a_{i}^{\dagger }$($v_{i}^{\dagger }$) operators forming the
evolution operator $\hat{L}$ to create the base states $\mid \vec{n}\rangle $
($\mid \vec{v}\rangle $) from a vacuum state $\mid 0\rangle $. Both types of
(independent from each other and hence commuting) operators fulfil the
relationship (for Paulions) 
\begin{equation}
a_{i}a_{j}^{\dagger }+a_{j}^{\dagger }a_{i}=\delta _{i,j}+2a_{j}^{\dagger
}a_{i}\left( 1-\delta _{i,j}\right) \mbox{.}  \label{CR}
\end{equation}
As mentioned above, the inherent properties of the FAM is the restriction of
the flip dynamics at cell $i$, $\sigma _{i}\leftrightarrow -\sigma _{i}$, 
\begin{equation}
\frac{1}{2}\sum_{j(i)}\langle n_{j}\mid \left( 1+\sigma _{j}\right) \mid
n_{j}\rangle =\sum_{j(i)}\langle n_{j}\mid \hat{A}_{j}\mid n_{j}\rangle \geq
f  \label{restriction}
\end{equation}
where $j(i)$ means a sum over all adjacent cells of $i$ and $f$ is the
restriction number. The number operators $\hat{A}_{j}$ and $\hat{V}_{j}$
denote $a_{j}^{\dagger }a_{j}$ and $v_{j}^{\dagger }v_{j}$ as usual.
Concerning the motion of information packages we postulate a diffusive
motion of the particles coupled to the existence of active cells at both
related sites. This exchange process due to Kawasaki enhances the mobility
of packages in active neighborhoods whereas it slows down inside a passive
cluster. Summarizing we obtain the evolution operator taking the form 
\begin{eqnarray}
L &=&L_{F}+L_{E}  \label{Liouville} \\
L_{F} &=&+\sum_{i}\lambda _{BA}(1-a_{i})a_{i}^{\dagger }\sum_{<m_{1}\cdots
m_{f},i>}\hat{A}_{m_{1}}\cdots \hat{A}_{m_{f}}  \nonumber \\
&&+\sum_{i}\lambda _{AB}(1-a_{i}^{\dagger })a_{i}\sum_{<m_{1}\cdots m_{f},i>}%
\hat{A}_{m_{1}}\cdots \hat{A}_{m_{f}}  \nonumber \\
L_{E} &=&+\sum_{<rs>}D_{0}\left[ v_{r}^{\dagger }v_{s}-\left( 1-\hat{V}%
_{r}\right) \hat{V}_{s}\right] \hat{A}_{r}\hat{A}_{s}+\mbox{symmetric term} 
\nonumber
\end{eqnarray}
with the quantities $\lambda _{AB}=\tilde{\lambda}\exp [h]$, $\lambda _{BA}=%
\tilde{\lambda}\exp [-h]$ and $D_{0}$ are the kinetic coefficients of the
diffusion process. They are appropriate thermodynamic weighted to fulfil the
detailed balance condition. The parameter $h$ corresponds to the inverse
temperature of a heat bath and describes the interconnectivity between the
cells. The higher $h$ is set the lower is the interconnectivity of the
network. The first term of $L_{F}$ reflects the flip process $\sigma
_{i}=-1\rightarrow \sigma _{i}=+1$ whereas the second term represents $%
\sigma _{i}=+1\rightarrow \sigma _{i}=-1$. The second part, $L_{E}$,
expresses the exchange process $V_{i}+V_{j}\leftrightarrow V_{j}+V_{i}$
related to the existence of an active link between adjacent cells. The term
in Eq.(\ref{Liouville}) 
\begin{equation}
\sum_{<m_{1}\cdots m_{f},i>}\hat{A}_{m_{1}}\cdots \hat{A}_{m_{f}}
\label{restriction2}
\end{equation}
represents the kinetic restriction which are mentioned above. The
abbreviation $<m_{1}\cdots m_{f},i>$ denotes the sets of all the $f$ lattice
cells neighbored to the cell $i$. The operator $\hat{A}_{m}$ yields a
non-zero result only if the cell $m$ is an active one, so that the
expression (\ref{restriction2}) differs from a zero value if it is applied
to a cell of interest surrounded by at least $f$ active cells.

\noindent Using Eq.(\ref{Average}) the temporal evolutions of the two
relevant observables $\langle \hat{A}_{k}\rangle $ and $\langle \hat{V}%
_{k}\rangle $ result in 
\begin{eqnarray}
\partial _{t}\langle \hat{A}_{k}\rangle  &=&\lambda _{BA}\sum_{<m_{1}\cdots
m_{f},k>}\langle \hat{B}_{k}\hat{A}_{m_{1}}\cdots \hat{A}_{m_{f}}\rangle
-\lambda _{AB}\sum_{<m_{1}\cdots m_{f},k>}\langle \hat{A}_{k}\hat{A}%
_{m_{1}}\cdots \hat{A}_{m_{f}}\rangle   \nonumber \\
\partial _{t}\langle \hat{V}_{k}\rangle  &=&2D\langle \bigtriangledown _{k}(%
\hat{A}_{k}^{2}\bigtriangledown _{k}\hat{V}_{k})\rangle 
\label{Evolution_Equation_V}
\end{eqnarray}
where we exploit the discrete form of the Laplacian 
\begin{equation}
\triangle _{k}O_{k}=\frac{1}{l^{2}}\sum_{r(k)}(O_{r}-O_{k})\mbox{.} 
\nonumber
\end{equation}
The diffusion coefficient is modified, $D=D_{0}l^{2}$, where $l$ is the
length of the lattice cell. The current for the diffusive motion of the
packages is given by 
\begin{equation}
j=-2D\hat{A}_{k}^{2}\bigtriangledown _{k}\hat{V}_{k}\mbox{.}  \label{Current}
\end{equation}
It is intuitively obvious that the effective diffusion coefficient should be
dependent on the squared concentration of the active cells, and therefore on
the number of active links.

\section{Mean-field solution}

\begin{figure}[h]
\epsfxsize=0.7\textwidth
\centerline{\epsffile{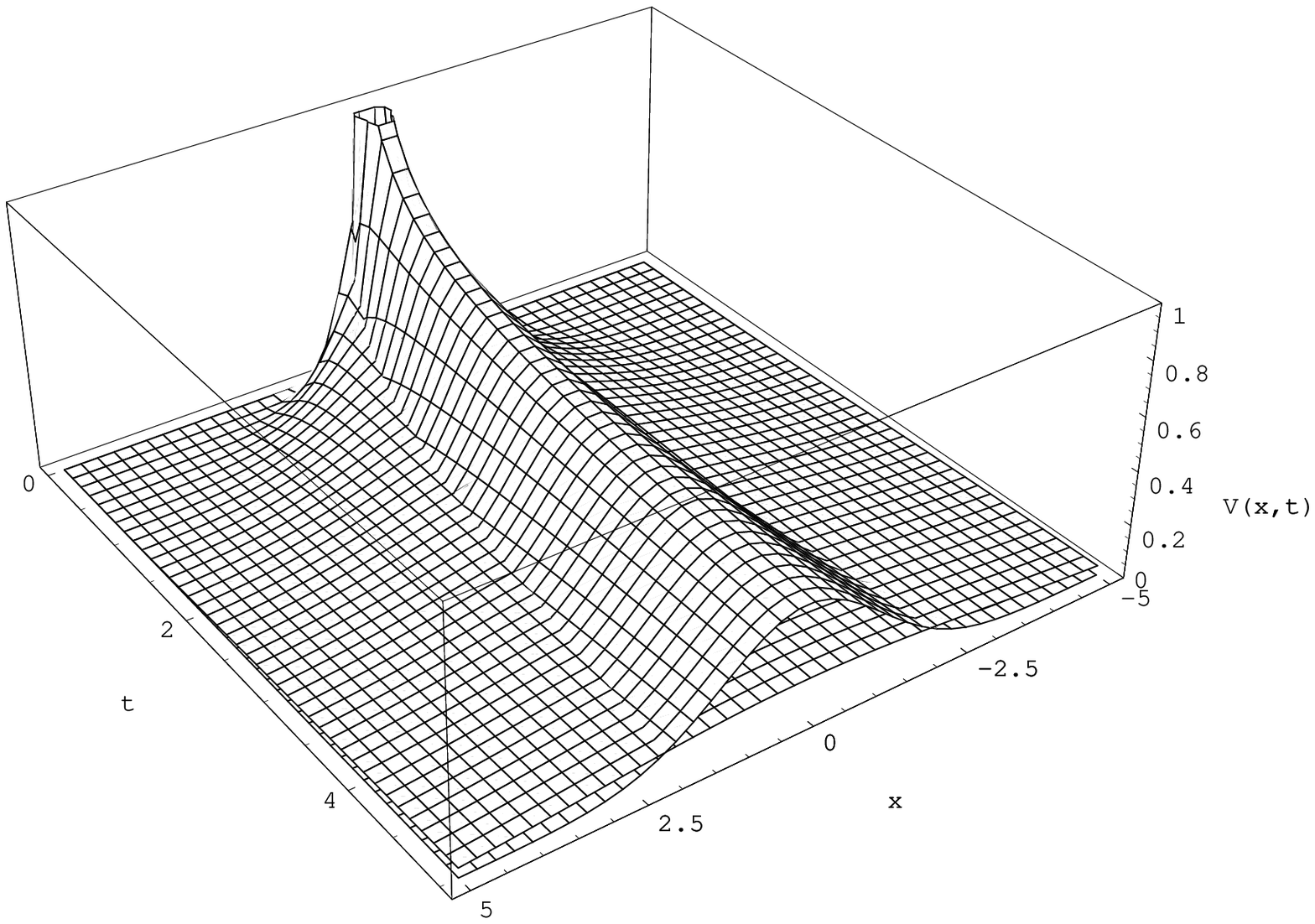}}
\caption{The diffusion of the packages concentration in the restricted
network model (upper ''localized'' graph) in comparison to the ordinary
diffusion process (lower graph) embedded in one dimension ($D=1$, $f=1$ and $%
A\left( 0\right) =\frac{1}{2}$).}
\label{Figure1}
\end{figure}

Because the formula for the vacancies concerns the same lattice index we may
neglect it. Within the next step we apply a mean-field approximation and
decouple all average values. Then, we obtain for the evolution equation of
the packages 
\begin{equation}
\partial _{t}\langle V\left( t\right) \rangle =2D\bigtriangledown \left[
\langle A\left( t\right) \rangle ^{2}\bigtriangledown \langle V\left(
t\right) \rangle \right]   \label{MFA-Vacancy}
\end{equation}
whereas the equation of motion for the mobile/active state yields 
\begin{equation}
\partial _{t}\langle A\left( t\right) \rangle =\lambda _{BA}\zeta \langle
1-A\left( t\right) \rangle \langle A\left( t\right) \rangle ^{f}-\lambda
_{AB}\zeta \langle A\left( t\right) \rangle ^{f+1}\mbox{.}
\label{MFA-Liquid}
\end{equation}
Its temporal solution is easily given by 
\begin{equation}
\langle A\left( t\right) \rangle =\bar{A}+\left[ A\left( 0\right) -\bar{A}%
\right] \exp \left( -\frac{t}{\tau _{1}}\right)   \label{Solution-Mobile}
\end{equation}
with the initial value $A\left( 0\right) $ and the steady state solution 
\begin{equation}
\bar{A}=\frac{\lambda _{BA}}{\lambda _{AB}+\lambda _{BA}}=\frac{1}{exp(2h)+1}%
\mbox{.}
\end{equation}
As the inverse relaxation time of the flip process we find 
\begin{equation}
\tau _{1}^{-1}=\left( \lambda _{BA}+\lambda _{AB}\right) \zeta \bar{A}^{f}
\label{Relaxation-time}
\end{equation}
with $\zeta =z\cdot \ldots \cdot (f+1)$. Notice that the steady state
solution in mean-field approximation is equal to the solution for the
paramagnetic lattice gas and independent of  $f$. In contrast, the
relaxation time depends on $\bar{A}^{f}$. Because the dynamics of the
packages is globally conserved the steady solution is fixed for all time by
the initial distribution, i.e. 
\begin{equation}
\bar{V}=\frac{\int V\left( x,0\right) dx}{\int dx}\mbox{.}
\end{equation}
Testing the stability we analyze the Eqs.(\ref{MFA-Vacancy}) and (\ref
{MFA-Liquid}) in a linear manner. Assuming $\delta A$ and $\delta V$ are
small fluctuations around the steady-state values we obtain by means of $%
\langle A\left( q,t\right) \rangle =\bar{A}+\delta A\left( \vec{q},t\right) $
and $\langle V\left( \vec{q},t\right) \rangle =\bar{V}+\delta V\left( \vec{q}%
,t\right) $ ($\vec{q}$ is the Fourier transformation of the local variable $%
\vec{x}$) 
\begin{equation}
\partial _{t}\left( 
\begin{array}{c}
\delta A\left( \vec{q},t\right)  \\ 
\delta V\left( \vec{q},t\right) 
\end{array}
\right) =-\left( 
\begin{array}{cc}
\tau _{1}^{-1} & 0 \\ 
0 & 2D\bar{A}^{2}q^{2}
\end{array}
\right) \left( 
\begin{array}{c}
\delta A\left( \vec{q},t\right)  \\ 
\delta V\left( \vec{q},t\right) 
\end{array}
\right) 
\end{equation}
The divergence of the second relaxation time $\tau _{2}^{-1}=2D\bar{A}%
^{2}q^{2}$ at the wave number $\vec{q}=\vec{0}$ reflects the global
conservation of the information in our network. Due to the quadratic
dependence on $\bar{A}$\ the relaxation time $\tau _{2}$ rapidly increases
if the active cells become passive. Obviously, the steady state reveals as
stable against pertubations indicated by the negative sign. To gain more
insight into the diffusion process of the information associated with
packages we consider its evolution equation (\ref{MFA-Vacancy}) applying the
mean-field solution for the active state (\ref{Solution-Mobile}). This leads
to a spatial independent, but temporal dependent effective diffusion
coefficient with 
\begin{equation}
\partial _{t}\langle V\left( t\right) \rangle =2D\left[ \bar{A}+\left(
A\left( 0\right) -\bar{A}\right) \exp \left( -\frac{t}{\tau _{1}}\right) 
\right] ^{2}\bigtriangledown ^{2}\langle V\left( t\right) \rangle \mbox{.}
\end{equation}
A measure for the fluctuations of the packages concentration is given by 
\begin{eqnarray}
F\left( t\right)  &=&\int_{0}^{t}\langle A\left( t%
{\acute{}}%
\right) \rangle ^{2}dt%
{\acute{}}%
\label{Solution-Vacancies} \\
&=&\bar{A}^{2}t+2\tau _{1}\bar{A}\left( A\left( 0\right) -\bar{A}\right)
\left( 1-e^{-\frac{t}{\tau _{1}}}\right) +\frac{\tau _{1}}{2}\left( A\left(
0\right) -\bar{A}\right) ^{2}\left( 1-e^{-\frac{2t}{\tau _{1}}}\right) %
\mbox{.}  \nonumber
\end{eqnarray}
Studying the asymptotic limits of $F\left( t\right) $ we recognize different
temporal regimes. Whereas for small times the fluctuations are dominated by
the initial value of the active cells $F\left( t\right) \sim A^{2}\left(
0\right) t$ the fluctuations are approximated by $F\left( t\right) \sim 
\bar{A}^{2}t$ for long times. Although the fluctuations go to infinity in
the long-time limit (in this mean-field theory) there is a
(quasi-)localization in an intermediate temporal range especially for small
interconnectivity (equivalent to low temperature). Only a small part of all
cells is active ($\bar{A}\ll \left( A\left( 0\right) -\bar{A}\right) $)
inhibiting the diffusion process. The exponential functions of the second
and the third term in Eq.(\ref{Solution-Vacancies}) are negligible so that
the fluctuations $F\left( t\right) $ remains almost constant. To see if
there is a quasi-localization one must compare the relaxation time $\tau $\
in Eq.(\ref{Relaxation-time}) and the time scale $t_{L}$ where the first
term is equal the second and the third term in Eq.(\ref{Solution-Vacancies}%
). In this connection, we assume that the exponential functions may be
neglected. The time $t_{L}$ associated with this point is (supposing $%
A\left( 0\right) >\bar{A}$) 
\begin{equation}
t_{L}=\frac{\tau _{1}}{2}\left[ \left( \frac{A\left( 0\right) }{\bar{A}}%
\right) ^{2}-1\right] \mbox{.}
\end{equation}
Obviously, a localization ($t_{L}\gg \tau _{1}$) becomes only possible if
the concentration $\bar{A}$ is small enough (sufficient low
interconnectivity/temperature) for the time interval $\tau _{1}<t<t_{L}$. A
rough estimation yields the ratio 
\begin{equation}
\frac{t_{L}}{\tau _{1}}\sim \exp \left( 4h\right) \mbox{.}
\end{equation}
Further, we may use $F\left( t\right) $ to transform the partial temporal
derivative of the evolution equation, 
\begin{equation}
\frac{\partial }{\partial t}=\frac{dF(t)}{dt}\frac{\partial }{\partial F(t)}%
=\langle A(t)\rangle ^{2}\frac{\partial }{\partial F(t)}\mbox{,}
\end{equation}
and hence to obtain the ordinary diffusion equation 
\[
\frac{\partial \langle V(t)\rangle }{\partial F(t)}=2D\bigtriangledown
^{2}\langle V(t)\rangle 
\]
applying the transformed Eq.(\ref{MFA-Vacancy}) 
\begin{equation}
\frac{\partial \langle V(t)\rangle }{\partial t}=\langle A(t)\rangle ^{2}%
\frac{\partial \langle V(t)\rangle }{\partial F(t)}=2D\langle A(t)\rangle
^{2}\bigtriangledown ^{2}\langle V(t)\rangle \mbox{.}
\end{equation}
If we now start with a $\delta $-distributed information (packages) density,
i.e. we drop down the information at one point and observe how the
information will be distributed in the time, 
\begin{equation}
V(\vec{x},0)=\bar{V}\delta \left( \vec{x}\right) 
\end{equation}
we get the result of the Eq.(\ref{MFA-Vacancy}) 
\begin{equation}
V(\vec{x},t)=\frac{1}{\left( 4\pi DF(t)\right) ^{\frac{d}{2}}}\exp \left( 
\frac{\vec{x}^{2}}{4\pi DF(t)}\right) \mbox{.}
\end{equation}
Hence, the behavior of $F(t)$ directly influences the diffusion of the
information. If $F(t)$ remains almost constant in a time regime the
diffusion process and therefore the distribution of the information stops,
and data packages are localized (compare Fig.\ref{Figure1}). 

\section{Conclusions}

In our extended facilitated kinetic Ising model we find a
(quasi-)localization due to the coupling of the diffusive dynamics for data
packages with the existence of the active cells at sufficiently low
temperatures. Because their part of all cells is small in this case so that
the diffusion almost stops. Therefore, the packages are fixed at or near the
initial position, information cannot be widespread. Thus, we may separate
two well-distinguished time regimes. First of all the fast flip (or
active-passive) relaxation takes place. The higher the restriction number $f$
the more the relaxation time slows down (compare Eq.(\ref{Relaxation-time}%
)). Then after a while of the relaxed system of active-passive states
sticking the packages the influence of the (slow) diffusion is remarkable,
equilibrates the packages concentration (information) of the total system.
However, for high enough temperature there are a sufficient part of active
cells so that the coupled diffusion can fasten the equilibration and
information can easily spread through the network. Like in \cite{Pigorsch1}
for the Fredrickson-Andersen model shown mean-field approximation may
provide false results. The mean-field solution leads to a dynamics which
completely breaks down below a critical (interconnectivity) temperature.
That means below a critical (interconnectivity) temperature information
would spread through a fixed network like through a sponge (Bond percolation
results would apply to this case, see e.g. \cite{Stauffer1}.). Here, we exploit
a more sophisticate approximation, taking local processes into account. But
we expect that the found quasi-localization alters in a permanent
localization at low temperature if  $f>z$. Because thus it exist stable
clusters of passive cells at any temperature in contrast to the case $f\leq z
$ where all cells can be activated \cite{Pigorsch1}. To prove this result
one may study the theorem of van Enter \cite{Enter} describing a bootstrap
percolation model \cite{Adler1} (or diffusion percolation\cite{Adler2}).

\end{document}